\documentstyle[12pt,epsf]{article}
\advance\textheight by \topskip
\begin{document}
\thispagestyle{empty}
\begin{flushright}
KUCP0155\\
August 1, 2001\\
\end{flushright}
\vskip 2 cm
\begin{center}
{\LARGE\bf Evolution of Cosmological Perturbations in the Brane World}
\vskip 1.7cm

 {\bf Kazuya Koyama}\footnote{
E-mail: kazuya@phys.h.kyoto-u.ac.jp} 
{\bf Jiro Soda}
\footnote{E-mail: jiro@phys.h.kyoto-u.ac.jp } \

\vskip 1.5mm

\vskip 2cm
 $^1$ Graduate School of Human and Environment Studies, Kyoto University, 
       Kyoto  606-8501, Japan \\
 $^2$ Department of Fundamental Sciences, FIHS, Kyoto University,
       Kyoto, 606-8501, Japan \\
\end{center}

{\centerline{\large\bf Abstract}}
\begin{quotation}
\vskip -0.4cm
The evolution of the cosmological perturbations is studied in the context of
the Randall-Sundrum brane world scenario, in which our universe is 
realized on a three-brane in the five dimensional Anti-de Sitter(AdS) 
spacetime.
We develop a formalism to solve the coupled dynamics of the 
cosmological perturbations in the brane world and the gravitational wave
in the AdS bulk. Using our formalism, the late time 
evolution of the cosmological scalar
perturbations at any scales larger than the AdS curvature scale
$l$ is shown to be identical with the
one obtained in the conventional 4D cosmology, provided the effect of
heavy graviton modes may be neglected. 
Here the late time means  the epoch when the Hubble horizon
$H^{-1}$ in the 4D brane world is sufficiently larger than the 
AdS curvature scale $l$. 
If the inflation occurs
sufficiently lower than $l^{-1}$, the scalar temperature anisotropies 
in the Cosmic Microwave Background at large scales can be 
calculated using the constancy of the
Bardeen parameter as is done in the 4D cosmology.
The assumption of the result is that 
the effect of the massive graviton with mass 
$m e^{-\alpha_0}>l^{-1}$ in the brane world is negligible,
where $e^{\alpha_0}$ is the scale factor of the brane world.
We also discuss the effect of these massive gravitons on the evolution
of the perturbations.
\end{quotation}
 \newpage

\def\k{\mbox{\boldmath$k$}}
\def\x{\mbox{\boldmath$x$}}
\baselineskip=0.6cm
\section{Introduction and Summary}
\setcounter{equation}0
Much attention has been paid for the possibility that we are living on a
3-brane in higher dimensional spacetime \cite{BW1,BW2}. 
This brane world picture alters the conventional notion of the extra 
dimensions. Particularly if the bulk is Anti de Sitter (AdS) spacetime,
the extra dimensions could be large or even infinite.
The action describing the brane world picture is given by
\begin{equation}
S= \frac{1}{2 \kappa^2}\int d^5 x \sqrt{-g}
\left(
{\cal R}^5 +  \frac{12}{l^2} \right)
- \sigma \int d^4 x \sqrt{-g_{brane}}
+ \int d^4 x \sqrt{-g_{brane}} {\cal L}_{matter},
\label{0-1}
\end{equation}
where ${\cal R}^5$ is the 5D Ricci scalar, $l$ is the curvature radius
of the AdS spacetime and $\kappa^2=8 \pi G$ 
where $G$ is the Newton constant in the 5D spacetime. 
The brane has tension $\sigma$ and the induced metric on the brane 
is denoted as $g_{brane}$. Matter is 
confined to the 4D brane world and is described by the
Lagrangian ${\cal L}_{matter}$. We will assume
$Z_2$ symmetry across the brane. 

Recently, Randall and Sundrum (RS) constructed a simple model for a brane
world \cite{RS}. 
They assumed the effect of the matter confined to the brane is
negligible compared with that of the surface tension. Their solution is
described by the metric;
\begin{equation} 
ds^2= \left( \frac{l}{z} \right)^2
\left( dz^2 -d \tau^2 + \delta_{ij} dx^i dx^j \right).
\label{0-2}
\end{equation}
It has been shown that the usual 4D gravitational interactions are
recovered on the 3-brane. 
One of the fascinating feature of their model is that 
the 5D spacetime is not necessarily compactified.

In RS model, the 3-brane is Minkowski spacetime. 
Solutions for homogeneous expanding brane world are
obtained by many people [4-14]. 
It has been shown that the evolution of the
universe is identical with that of the conventional 4D cosmology 
at sufficiently low energies. However in the real world, the universe has
inhomogeneity which leads to our structure of the universe [15-17]. 
This inhomogeneity can be observed today, for example, in the Cosmic
Microwave Background Radiation (CMB). 
Then the cosmological perturbations in the
brane world give direct tests for a viability of the brane world
idea. In addition, the inhomogeneous fluctuations on the brane could be
a powerful observable to probe the existence of the extra dimensions. This is
because the inhomogeneous fluctuations on the brane inevitably
produce the perturbations of the bulk geometry \cite{SMS}. 
The perturbations 
in the bulk affect the motion of the brane in turn. Then, in general, 
the dynamics on the brane cannot be separated from the dynamics 
in the bulk.  This could add a new property to the evolution of the
cosmological perturbations and could reject the brane world idea. 

To study the evolution of cosmological perturbations, we should 
treat the coupled system of brane-bulk dynamics. The problem has
the similarity with the dynamics of the domain wall interacting 
with the gravitational wave, 
which has been investigated in 4D spacetime \cite{II}. 
In our case, the matter on the brane is dynamical. This makes 
the problem very difficult. We should find a solution for the
brane with the cosmological expansion and
inhomogeneous fluctuations. 
The most straightforward way is to solve the 5D Einstein equation, 
however, it would be difficult to carry out in general.

In this paper we propose a new method to
deal with the problem.  We observe that 
the brane world cosmology can be constructed
by cutting the perturbed AdS spacetime along the suitable slicing and 
gluing two copies of remaining spacetime. The point is as follows.
If we choose a slicing to cut the 5D AdS spacetime, the 
jump of the extrinsic curvature along the slicing is determined. 
Because the jump of the extrinsic curvature should be equated with the
matter localized on the brane, the matter on the brane is also
determined. In other words, a solution for a brane with the given matter
can be obtained by finding a suitable slicing. 
To find the suitable slicing for the given matter,  
we need two kinds of coordinate transformations.
One is a large coordinate transformation which leads to the 
slicing which determines the background matter. Another is an infinitesimal
coordinate transformation which leads to the slicing which determines
matter perturbations. The coordinate
transformations will be determined by imposing the conditions on the 
matter such as
equation of state. More detailed procedures 
will be described in the next section.

Our main result is 
\begin{equation}
\frac{\delta \rho}{\rho}=-2 \Phi_0 =const.
\label{ee-0}
\end{equation}
at superhorizon scales and for late times
when the Hubble scale $H$ is sufficiently low ($H \ll l^{-1}$).
Here $\delta \rho$ is the density fluctuations and 
$\Phi_0$ is the metric perturbations 
in the longitudinal gauge in the brane world
and we assumed the barotropic index of the matter is
constant. The point to observe is that the solution (\ref{ee-0})
is identical with the one obtained in the 4D cosmology.
We can also show that the late time evolution of the perturbations 
agrees with the one obtained in the 4D cosmology
at subhorizon scales larger than the AdS curvature scale $l$.

The assumption to obtain the above results is that 
the effect of the massive graviton with mass 
$m > m_{eff}=l^{-1} e^{\alpha_0}$ 
in the brane world is negligible
where $e^{\alpha_0}$ is the scale factor of the brane world. 
We can understand the fact that 
the massive graviton with $m > m_{eff}$ can modify
the evolution from the following 
arguments. For late times, the cosmological brane approaches to the RS brane. 
For the RS brane, the 4D gravity 
is recovered by the zero-mode of 5D graviton [3,20-23]. 
The Kaluza-Klein modes
give the correction to the 4D gravity. However, in the Anti-de 
Sitter spacetime, the brane is protected from the Kaluza-Klein modes
by the potential barrier which arises from the curvature of the AdS 
spacetiem. For earlier times, 
the cosmological brane is located at larger $z$ in the
coordinate (\ref{0-2}). The point is that, for larger $z$, 
the potential barrier becomes lower. 
Then the relatively light graviton can interact with the 
brane. Thus for early times, the 4D cosmology will be susceptible
to the Kaluza-Klein modes of 5D graviton. If the brane interacts with 
the 5D gravitational perturbations, the gravitational waves are inevitably 
emitted to the 5D bulk.
It will cause the modification in the evolution of the perturbations.
This picture 
is consistent with the result that the modes with large $m > m_{eff}
=l^{-1} e^{\alpha_0}$ can modify the evolution 
because $m_{eff}$ becomes smaller for earlier times.

The paper is organized as follows. In section II,
we describe our formalism in details and derive the
background solution using it. 
In section III, we calculate the perturbations 
at superhorizon scales for late times $H < l^{-1}$
using the formalism. 
In  section VI, we calculate the perturbations 
at subhorizon scales for late times $H < l^{-1}$. 
It will be shown that the evolution of the perturbations
is identical with the one obtained in the 4D cosmology
for any scales larger than the AdS curvature scale, if
the effect of the massive graviton $m > m_{eff}$ is
negligible. In section V we study the effect of the
massive graviton $m > m_{eff}$ on the evolution of the
perturbations.
Finally we discuss the implication of our results on
the brane world cosmology. In the Appendix, we listed useful
formulas for calculations.

\section{Formalism}
\setcounter{equation}0
\subsection{Background}
We shall start with the Randall and Sundrum solution for 
the barne world \cite{RS}.
They considered a single brane with positive tension $\sigma$ in the
5D anti-de Sitter spacetime.  
Setting the surface tension of the brane by
\begin{equation}
\kappa^2 \sigma=\frac{6}{l},
\label{1-1}
\end{equation}
and assuming the $Z_2$ symmetry across the brane, 
they found a solution described by the metric ({\ref{0-2}}).
The brane is located at $z=l$ (see Fig.1). 
From the metric (\ref{0-2}), we see the brane world is Minkowski 
spacetime.

\begin{figure}[ht]
  \epsfysize=80mm
\begin{center}
  \epsfbox{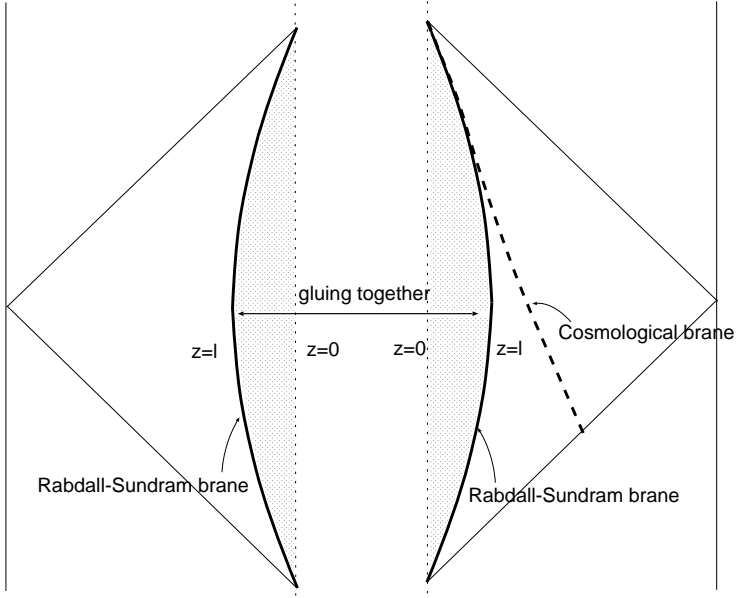}
\end{center}
  \caption{Conformal diagram of the AdS spacetime. The thick line 
shows the trajectory of the RS brane and the dotted line shows
the brane with cosmological expansion. The RS solution is obtained by
deleting the AdS spacetime from $z=l$ to the boundary $z=0$ (shaded
region) and gluing two copies of the remaining spacetime.}
  \label{fig1}
\end{figure}
           
Next we will seek the brane world with the cosmological 
expansion. For this purpose, we note that
the RS solution can be obtained by the following
procedure. First cut the AdS spacetime along $z=l$ and delete the
AdS spacetime from $z=l$ to the boundary $z=0$. Next glue two copies of the 
remaining spacetime along $z=l$ (see Fig.1). 
The jump of the extrinsic curvature
at $z=l$ should be equated with the matter at $z=l$.
Thus we must put the 
suitable matter on the brane to glue the spacetime. 
Then the content of the matter on the brane is restricted as (\ref{1-1}).
The above argument implies that if we use the different slicing to cut
the AdS spacetime, we need different matter to glue the spacetime. 
This is because the jump of the extrinsic curvature depends on 
the slicing we use to cut the AdS spacetime. 
Thus if we can find appropriate slicing to cut the AdS spacetime,
we can put suitable matter resulting the cosmological expansion 
on the brane (see Fig.2).

\begin{figure}[ht]
  \epsfysize=80mm
\begin{center}
  \epsfbox{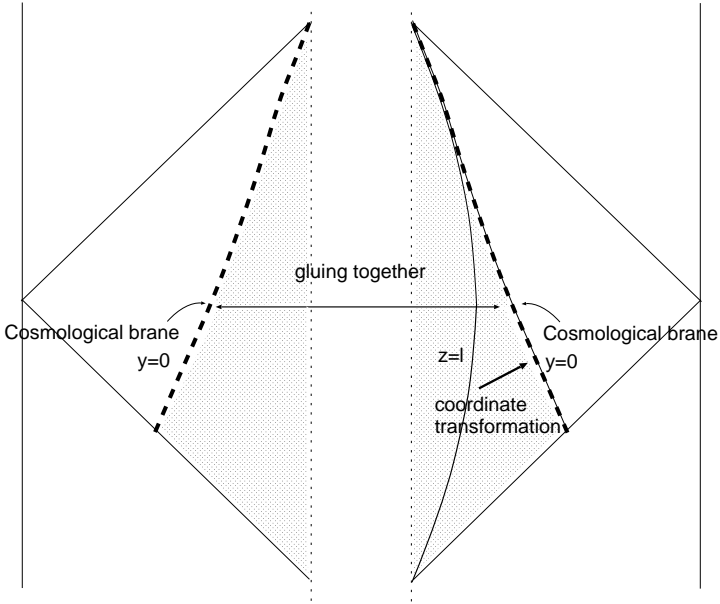}
\end{center}
  \caption{The solution for a brane with cosmological expansion 
is obtained by deleting the AdS spacetime from $y=0$ to the boundary 
(shaded region) and gluing two copies of the remaining spacetime.}
  \label{fig2}
\end{figure}
Now, we explicitly carry out the procedure to find the appropriate slicing. 

\hspace{1cm}\\
(1)Start with the AdS spacetime;\\
\begin{equation}
ds^2= \left( \frac{l}{z} \right)^2
\left( dz^2 -d \tau^2 + \delta_{ij} dx^i dx^j \right).
\label{1-2-1}
\end{equation}

\hspace{1cm}\\
(2)Make the coordinate transformation from the coordinate system
$(z, \tau ,x^i) $ to $(y,t,x^i)$ by 
\begin{equation}
  z=l \left(f(u)-g(v)\right), \quad \tau=l \left(f(u)+g(v) \right),
\label{1-3}
\end{equation}
where $u,v$ are the null coordinates of the new coordinate system;
$u=(t-y)/l,v=(t+y)/l$ and $f(u)$ and $g(v)$ are the arbitrary functions.
The resulting metric is 
\begin{equation}
ds^2 =  4 \frac{f'(u)g'(v)}{(f(u)-g(v))^2}(dy^2-dt^2)
+\frac{1}{(f(u)-g(v))^2} \delta_{ij} dx^i dx^j, \\
\label{1-4}
\end{equation}
where $f'(x)=d f(x)/d x$ and $g'(x)=d g(x)/d x $.
For the future convenience, we put
\begin{equation}
e^{2 \beta(y,t)} =  4 \frac{f'(u)g'(v)}{(f(u)-g(v))^2}, \quad
e^{2 \alpha(y,t)} = \frac{1}{(f(u)-g(v))^2}.
\label{1-5}
\end{equation}

\hspace{1cm}\\
(3)Delete the AdS spacetime from $y=0$ to
the boundary and glue two copies of remaining spcaetime. 
The jump of the extrinsic curvature 
at the brane ($y=0$) is determined by the first derivative of the metric with 
respect to $y$;
\begin{eqnarray}
\alpha_1 (t) &=& \frac{1}{l} 
\frac{f'(t/l)+g'(t/l)}{f(t/l)-g(t/l)}, \nonumber\\
\beta_1 (t) &=&  \frac{1}{l} \left( 
\frac{f'(t/l)+g'(t/l)}{f(t/l)-g(t/l)}+\frac{1}{2}
\frac{-f''(t/l)g'(t/l)+f'(t/l)g''(t/l)}{f'(t/l)g'(t/l)} \right),
\label{1-6}
\end{eqnarray}
where we denote the power series expansion near the brane as
\begin{equation}
\alpha(y,t)=\alpha_0(t)+ \alpha_1(t) \vert y \vert  + 
\frac{\alpha_2(t)}{2} y^2 
\cdot \cdot \cdot.
\label{1-6-1}
\end{equation}
The jump of the extrinsic curvature should be equated with
matter on the brane.  
Taking the 5D energy momentum tensor as 
\begin{equation}
T^M_N= diag(0,-\rho,p,p,p) \delta(y), 
\label{1-8}
\end{equation}
the junction condition can be read off as (see Appendix A)
\begin{eqnarray}
\alpha_1(t) &=& - \kappa^2 e^{\beta_0} \left(\frac{\sigma}{6}   
+\frac{\rho(t)}{6} \right), \nonumber\\
\beta_1(t) &=&  - \kappa^2 e^{\beta_0} 
\left(\frac{\sigma}{6}-\frac{\rho(t)}{3}-\frac{p(t)}{2} \right).
\label{1-9}
\end{eqnarray}

\hspace{1cm}\\
(4)Determine the matter content on the brane by imposing the equation of state 
\begin{equation}
p=w \rho.
\label{1-13}
\end{equation}
Then it gives one constraint on $f$ and $g$.
There remains one freedom in $f$ and $g$.
Since $e^{\beta_0(y=0,t)}$ determines
the time slicing in the brane world, it is a gauge freedom
in the brane world. We fix the gauge degree of freedom by demanding that
$t$ is the cosmological time;
\begin{equation}
e^{2 \beta_0}=  4 \frac{f'(t/l)g'(t/l)}{(f(t/l)-g(t/l))^2}=1.
\label{1-14}
\end{equation}
Combining (\ref{1-13}) and (\ref{1-14}),
we can determine the function $f(t)$ and $g(t)$. 
The $y$ dependence of the metric can be obtained
automatically by replacing $f(t)$ to $f(u)$ and $g(t)$ to $g(v)$.
Hence we obtained the coordinate transformation which leads to
a brane with matter of given equation of state. 

The induced metric on the brane becomes
\begin{equation}
ds^2 =-dt^2+e^{2 \alpha_0(t)} \delta_{ij} dx^i dx^j,
\label{1-15}
\end{equation}
where $e^{\alpha_0(t)} =(f(t/l)-g(t/l))^{-1}$ 
is the scale factor of the brane world.
From (\ref{1-13}) and (\ref{1-14}), we can verify that 
$\alpha_0$ and $\rho$ satisfy 
\begin{eqnarray}
&& \dot{\rho} + 3 \dot{\alpha_0}(\rho+p)=0, \nonumber\\
&& \dot{\alpha_0}^2 = 
\frac{8 \pi G_4}{3}\rho+ \frac{\kappa^4 \rho^2}{36}, 
\label{1-16}
\end{eqnarray}
where $\kappa^4 \sigma=48 \pi G_4$. The former is the usual 
conservation of the energy. Since the term proportional to 
$\rho^2$ falls rapidly, the latter is identical with the
Friedmann equation for late times. 
We show the solution of $f(u)$ and $g(v)$ for late times
in Appendix A. The solution has two constants of the
integration. We will normalize the scale factor as 
$e^{\alpha_0(t=t_{present})}=1$.

\subsection{Perturbations}
In the previous subsection, we obtained the brane world with
the cosmological expansion. In the real world the universe has inhomogeneity
which leads to our structure of the universe. Then it is important 
to obtain the solution for an inhomogeneous brane world.
In this paper, we will concentrate our attention to the
scalar perturbations. 
Unlike the homogeneous brane, we cannot place the inhomogeneous
brane in the exact AdS spacetime as is shown in Ref \cite{SMS}. 
This is because the inhomogeneous perturbations in the brane world
inevitably produces the perturbations in the geometry of 
the bulk. The perturbations in the bulk affects the motion of the 
brane in turn. Then the equations for metric perturbations 
and matter perturbations in the brane world
can not be separated from the dynamics in the bulk. 
We should solve the 5D perturbations at the same time. 
The coupled equations for the brane dynamics and gravitational perturbations 
in the bulk are in general very difficult to deal with. 

The non-separable nature of the brane-bulk dynamics can be seen from the
power series expansion of the 5D Einstein equation near the brane. 
We will denote the power series expansion near the brane as
in (\ref{1-6-1}).
The dynamical variables of the brane are the potential perturbations
$\Phi_0$, curvature perturbations $\Psi_0$, density perturbations
$\delta \rho$ and velocity perturbations $v$ (see Appendix B).
We have two equations from conservations of energy-momentum tensor
$T^{\mu \nu}_{ \:\:\:\:\: ; \mu}=0$ 
\begin{eqnarray}
\dot{\delta \rho} &=&(\rho+p)(3 \dot{\Psi}_0+ e^{-\alpha_0} \nabla^2 v)
-3 \dot{\alpha}_0 
\left( \delta \rho+ \delta p \right),\nonumber\\
\left( (\rho+p) e^{\alpha_0} v \right)^{\cdot}
&=& -3 \dot{\alpha}_0 (\rho+p) e^{\alpha_0} v + \delta p 
+(\rho+p) \Phi_0,
\label{ee-5}
\end{eqnarray}
and the trace part of the Einstein equation
in the brane world from 5D Einstein equation; 
\begin{eqnarray}
\ddot{\Psi}_0+4 \dot{\alpha}_0 \dot{\Psi}_0 +\dot{\alpha}_0 \dot{\Phi}_0
+2(\ddot{\alpha}_0+2 \dot{\alpha}_0^2) \Phi_0 
- \frac{1}{3} e^{-2 \alpha_0} (2 \nabla^2 \Psi_0 -\nabla^2 \Phi_0)
\nonumber\\
=
\frac{\kappa^2}{3} \left(\frac{\beta_1}{2} 
\delta \rho -\frac{3 \alpha_1}{2} \delta p \right). 
\label{ee-6}
\end{eqnarray}
In the conventional 4D cosmology, in addition to these equations, 
we have the equation 
\begin{equation}
\Phi_0-\Psi_0 = 0,
\end{equation}
for matter with no anisotropic stress. 
Then we have closed set of the equations.
However in the brane world, the correspondent equation derived 
from the 5D Einstein equation is
\begin{equation}
E_2= - e^{-2 \alpha_0}(\Phi_0-\Psi_0+N_0),
\end{equation}
where $E$ is the non-diagonal $(i,j)$ comopnent and $N$ is the
$(y,y)$ component of the metric perturbations.
The equation contains $E_2$, so we cannot have
closed set of equations for the variables on the brane. 
This is because 
the inhomogeneous fluctuations on the brane inevitably produces 
the gravitational wave in the bulk, which gives the effective 
anistoropic stress to the perturbations. 

However, the procedure to obtain the background solution can be
applied for the inhomogeneous brane. The strategy is as follows. 
We first consider the perturbed 5D AdS spacetime 
in the coordinate system (\ref{1-2-1}). We assume the 
perturbations are small enough to treat by linear perturbations. 
Because there is no matter in the bulk,
the perturbations should satisfy the vacuum wave equation in the bulk. 
In the coordinate system (\ref{0-2}),
the wave equation can be solved easily 
with the help of the transverse-traceless (TT) gauge.
In the 5D spacetime, the free graviton has five independent components
which include one scalar component. Thus there is one variable
for the choice of the perturbed AdS spacetime. 
Next, we take the coordinate transformation (\ref{1-3})
to provide the cosmological background.
The transformation function $f(u)$ and $g(v)$ is determined by the 
background matter. The perturbations in the coordinate system $(y,t,x^i)$
is then easily obtained by the usual procedure of the coordinate
transformation.

Once the perturbed AdS spacetime is obtained, 
one might attempt to cut the spacetime along $y=0$ and glue 
two copies of remaining spacetime as is done for the background spacetime. 
However, we need to be more careful. 
The presence of matter on the brane bends the
brane. For the background matter we made coordinate transformation 
so that the brane is located at $y=0$. However
the matter perturbations also bends the brane. Then the perturbed brane
is no longer located at $y=0$ (see Fig.3). 
The perturbations evaluated at $y=0$ is {\it not} the perturbations 
induced on the barne.
Since the observers in the brane world are confined to the brane,
we should evaluate the perturbations induced on the brane. 
 Thus we should make (infinitesimal) 
coordinate transformation $\bar{x}^M = x^M + \xi^M$ to ensure that 
$\bar{y}=0$ denotes the location of the brane.
In general, the coordinate transformation
makes $g_{y \mu} (\mu=t,x^i)$ nonzero.
These components can be gauged away,
so we will take $g_{y \mu}=0$. 
Now the gluing can be performed so as to determine the cosmological
perturbations.

Then we impose the conditions on the matter perturbations to 
determine the variable for the choice of the perturbed AdS spacetime
and the infinitesimal coordinate transformation. 
The point is that imposing the conditions on the matter is equivalent
to solving the Einstein equation. Let us remember that 
we impose the equation of state $p=w \rho$ (\ref{1-13}) on the matter 
in deriving the background solution, which gives the Friedmann
equation and conservation of the energy on the brane (\ref{1-16}). 
The same is true of the perturbations.

\begin{figure}[t]
  \epsfysize=70mm
\begin{center}
  \epsfbox{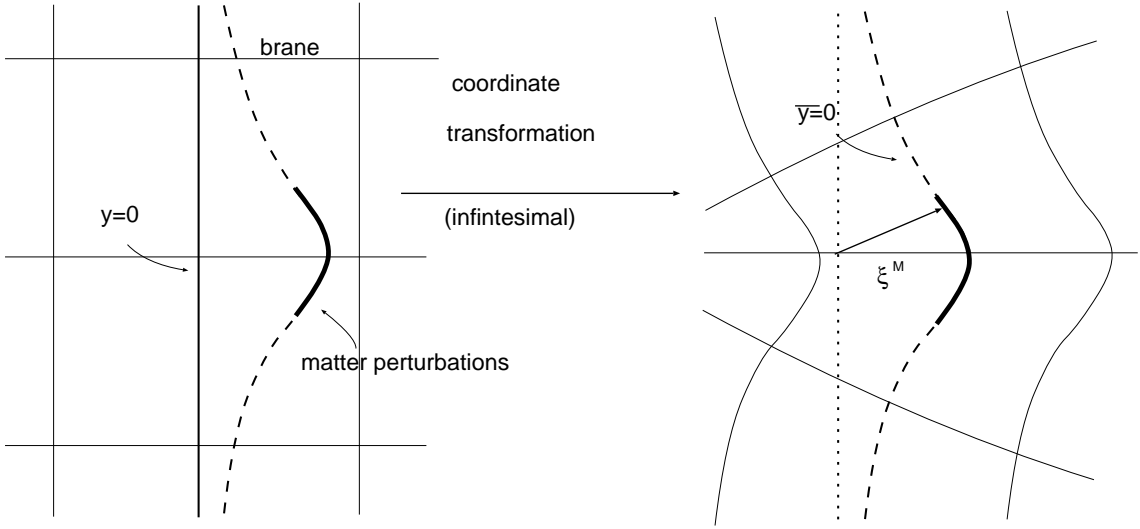}
\end{center}
  \caption{Bending of the brane due to the matter perturbations.
Due to the bending, the brane is not located at $y=0$. Furthermore, 
the brane is not perpendicular to the $y$ axis. We should make 
infinitesimal coordinate transformation.}  
\end{figure}

We summarize the procedure to obtain the inhomogeneous brane.

\hspace{1cm}\\
(1) Let us start with the AdS spacetime 
with linear scalar perturbations;
\begin{equation}
ds^2= \left(\frac{l}{z} \right)^2
\left(dz^2 - (1+2 \phi) d \tau^2 
+2 b_{,i} dx^i d \tau
+ \left((1 - 2 \hat{\Psi}) \delta_{ij}+ 
2 \hat{E}_{,ij} \right)dx^i dx^j\right), 
\label{2-B-1}
\end{equation}
where $b_{,i}$ denotes $\partial b/\partial x^i$ and $E_{,ij}$ denotes
$\partial^2 E/\partial x^i \partial x^j$.
The independent component of the scalar 5D graviton is one.
We will use the transverse-traceless gauge to fix the gauge freedom.
The gauge fixing conditions are given by
\begin{eqnarray}
&& \phi-3 \hat{\Psi} + \nabla^2 \hat{E} = 0, \nonumber\\
&& 2 \dot{\phi} + \nabla^2 b =0, \nonumber\\
&& \dot{b}+ 2 \hat{\Psi} -2 \nabla^2\hat{E}=0. 
\label{2-B-2}
\end{eqnarray}
Using these conditions, the Einstein equation in the bulk becomes
\begin{equation}
\frac{\partial^2 h}{\partial z^2} -\frac{3}{z} \frac{\partial
h}{\partial z}-\frac{\partial^2 h}{\partial \tau^2}+ \nabla^2 h=0,
\label{B-2-3}
\end{equation}
where $h=\phi, b, \hat{\Psi}$ and $\hat{E}$.
Taking the solutions of the form 
$h(z,\tau,x^i)=(z/l)^2 S(z)  e^{-i \omega \tau} e^{i \k \x} $, 
the equation for $S(z)$ is obtained as
\begin{equation}
S(z)''+ \frac{1}{z} 
S(z)' +\left(m^2 -\frac{4}{z^2} \right)S(z)=0, 
\:\: m^2=\omega^2-\k^2,
\label{B-2-4}
\end{equation}
where $S(z)'$ denotes $d S(z)/dz$.
The solutions are given by linear combinations of the Bessel function
and the Neumann function $J_2(mz)+ a_m N_2(mz)$. The 
coefficient $a_m$ is determined by the 
the boundary conditions at $z \to \infty$. 
We take the boundary conditions that the positive
frequency functions are ingoing at $z \to \infty$. Then the solution
is given by
\begin{eqnarray}
\hat{h} = \left( \frac{z}{l}\right)^2 \int \frac{d^3 \k}{(2 \pi)^3}
\int d m \:\:
h_{m}(\k)  H^{(1)}_2 (m z) e^{-i \omega \tau}e^{i \k \x}, 
(\hat{h}= \phi,b,\hat{\Psi},\hat{E}),
\label{B-2-5}
\end{eqnarray}
where $H^{(1)}_2$ is the Hunkel function of the first kind. 
From the gauge fixing conditions, the coefficients $h_{m}(\k)$ 
satisfy 
\begin{eqnarray}
\phi_{m}(\k) &=& 
\frac{2 \k^4 l^2}{2 \k^2+ 3 m^2}  E_{m}(\k), \nonumber\\
b_{m}(\k) 
&=& -\frac{4 i \sqrt{\k^2+m^2}\k^2 l^2}{2 \k^2 +3 m^2} 
E_{m} (\k), \nonumber\\
\hat{\Psi}_{m}(\k) &=& 
-\frac{m^2 \k^2 l^2}{2 \k^2 +3 m^2} E_{m}(\k), \nonumber\\
\hat{E}_{m}(\k)&=& l^2 E_{m}(\k),
\label{B-2-6}
\end{eqnarray}
where $E_{m}(\k)$ is the arbitrary coefficient. This corresponds to the
one degree of freedom of the 5D scalar graviton and 
represents the spectrum of the
gravitational waves emitted from the perturbed brane. 

\hspace{1cm}\\
(2) Make the coordinate transformation to provide the cosmological
background;
\begin{equation}
  z=l \left(f(u)-g(v)\right), \quad \tau=l \left(f(u)+g(v) \right),
\label{B-2-7}
\end{equation}
The pertrubations in the cosmological background can be obtained 
using (\ref{1-5}) as
\begin{eqnarray}
h = J \:\: e^{-2\alpha(y,t)} \int \frac{d^3 \k}{(2 \pi)^3}
\int d m \:\:
h_{m}(\k) H^{(1)}_2(ml e^{-\alpha(y,t)}) 
e^{-i \omega \tau(t,y)} e^{i \k \x}, 
\label{B-2-8}
\end{eqnarray}
where $J$ is the factor which comes from the Jacobian of the
transformation (\ref{B-2-7});
\begin{eqnarray}
\frac{\partial \tau}{\partial y} 
&=& l(-f'(u)+g'(v))=
l \dot{\alpha} e^{-\alpha},\:\:\:\:  
\frac{\partial z}{\partial y} =-l (f'(u)+g'(v))=
-l \alpha' e^{-\alpha}, \nonumber\\
\frac{\partial \tau}{\partial t}
&=& l(f'(u)+g'(v))=
l \alpha'  e^{-\alpha},\quad \:\: 
\frac{\partial z}{\partial t} 
= l(f'(u)-g'(v))=
-l \dot{\alpha}  e^{-\alpha}.
\label{B-2-9}
\end{eqnarray}
In addition, we make (infinitesimal) coordinate transformations 
$\bar{x}^M=x^M+\xi^M$ to ensure that the brane is located at
$\bar{y}=0$. 
After imposing the gauge conditions $g_{y\mu}=0$, there remains one freedom of 
the coordinate transformation ($\xi^y$).
Then we take a slicing along the spacetime $\bar{y}=0$ to 
cut the spacetime. 

\hspace{1cm}\\
(3) Cut the spacetime at $\bar{y}=0$ and 
glue two copies of the remaining spcaetime along $\bar{y}=0$.
From the junction conditions the matter on the brane is determined
in terms of $E_{m}(\k)$ and $\xi^y$.

\hspace{1cm}\\
(4)Finally impose the two conditions on the matter 
perturbations and  determine $E_{m}(\k)$ and $\xi^y$.
We will impose the condition on the anisotropic stress
and equation of state of the matter.

%%%%%%%%%%%%%%%%%%%%%%%%%%%%%%%%%%%%%%%%%%%%%%%%%%%%%%%%%%
%%%%%%%%%%%%%%%%%%%%%%%%%%%%%%%%%%%%%%%%%%%%%%%%%%%%%%%%%%
%%%%%%%%%%%%%%%%%%%%%%%%%%%%%%%%%%%%%%%%%%%%%%%%%%%%%%%%%%
%%%%%%%%%%%%%%%%%%%%%%%%%%%%%%%%%%%%%%%%%%%%%%%%%%%%%%%%%%

\section{Evolution of Perturbations at Superhorizon Scales}
\setcounter{equation}0
Following the formalism developed in the previous section,
we calculate the evolution of the perturbations.
To simplify the calculations, we first consider the long-wave 
perturbations in the brane world. We shall take the limit 
\begin{equation}
\k \to 0.
\label{3-a-1}
\end{equation}
We will calculate the evolution for late times where 
the Hubble scales is sufficiently low
\begin{equation}
H \ll l^{-1}.
\end{equation}
We will take the assumption that the modes with 
$m e^{-\alpha_0}> l^{-1}$ do not contribute to the
perturbations in the brane world. Then we assume
\begin{equation}
ml e^{- \alpha_0} \ll 1.
\end{equation}
The effect of these modes will be discussed in the
section V.

\subsection{Calculation of Perturbations}

\hspace{1cm} \\
(1) In the $\k \to 0$ limit, we see only $\hat{E}$ survives
in (\ref{B-2-6}). 
Then we shall start with 
\begin{equation}
ds^2= \left(\frac{l}{z} \right)^2
\left(dz^2 - d \tau^2 
+ \left(\delta_{ij}+ 2 \hat{E}_{,ij} \right)dx^i dx^j\right). 
\label{3-b-2}
\end{equation} 

\hspace{1cm}\\
(2)We make the (large) coordinate transformation (\ref{1-3}).
Since $\hat{E}$ does not change in this coordinate 
transformation, the metric is given by
\begin{equation}
ds^2=e^{2 \beta(y,t)} \left( dy^2 -d t^2 \right)+  e^{2 \alpha(y,t)}
\left(\delta_{ij}+ 2 \hat{E}_{,ij}\right) dx^i dx^j.
\label{3-b-4}
\end{equation}
Due to the bending of the brane by matter perturbations,
the brane is not located at $y=0$. 
We perform the (infinitesimal) coordinate transformation 
\begin{equation}
x^M \to x^M + \xi^M, \quad \xi^M=(\xi^y,\xi^t,\xi^{,i}).
\label{3-b-5}
\end{equation}
After this coordinate transformation, the perturbed metric is given by
\begin{eqnarray}
ds^2 &=& e^{2 \beta(y,t)} \left((1+2 N) dy^2 -(1+2 \Phi)
d t^2 +2 A \: dt \: dy \right) \nonumber\\
 && +  e^{2 \alpha(y,t)}
\left(\left((1 - 2 \Psi) \delta_{ij}+ 2 E_{,ij}\right) 
dx^i dx^j+2 B_{,i} dx^i d t + 2 G_{,i} dx^i dy \right),
\label{3-b-6}
\end{eqnarray}
where
\begin{eqnarray}
\Phi &=& \dot{\xi}^t+ \beta' \xi^y + \dot{\beta} \xi^t, \nonumber\\
\Psi &=& - \dot{\alpha} \xi^t -\alpha' \xi^y, \nonumber\\
E &=& \hat{E} + \xi, \nonumber\\
B &=& \dot{\xi} - e^{2 (\beta-\alpha)} \xi^t, \nonumber\\
A &=& \dot{\xi}^y -\xi^{t'}, \nonumber\\
G &=& e^{2(\beta-\alpha)} \xi^y+\xi', \nonumber\\
N &=& \xi^{y'}+ \beta' \xi^y +\dot{\beta} \xi^t.
\label{3-b-7}
\end{eqnarray}          
We will denote the power series expansion near the brane as
\begin{equation}
\Phi(y,t)=\Phi_0(t)+ \Phi_1(t) \vert y \vert + \cdot \cdot \cdot.
\end{equation}
We take the gauge condition $G=A=0$ and $B_0=0,E_0=0$ 
(see appendix B-1). 
This determines $\xi^t$ and $\xi$ in terms of $\xi^y$;
\begin{eqnarray}
\xi^t &=& \int^y_0 dy \dot{\xi}^y + \hat{T}_0, \quad
\hat{T}_0=- e^{2 \alpha_0} \dot{\hat{E}}_0 , 
\nonumber\\
\xi &=& -\int^y_0 dy e^{2(\beta-\alpha)} \xi^y -\hat{E}_0.
\label{3-b-8}
\end{eqnarray}
Then we obtain the metric perturbations induced on the brane
\begin{eqnarray}
\Phi_0 &=& \beta_1 \xi^y_0 + \dot{\hat{T}}_0 ,\nonumber\\
\Psi_0 &=& - \alpha_1 \xi_0^y  - \dot{\alpha}_0 \hat{T}_0  \nonumber\\
N_0 &=& \xi^y_1 + \beta_1 \xi^y_0,
\label{3-b-9}
\end{eqnarray}
and the first derivative of the metric perturbations 
\begin{eqnarray}
\Phi_1 &=& \ddot{\xi}^y_0 + \beta_1 \xi^{y}_1+ \beta_2 \xi_0^y +
\dot{\beta}_1 \hat{T}_0, \nonumber\\
\Psi_1 &=& 
- \alpha_1 \xi^{y}_1-\dot{\alpha}_0 \dot{\xi}^y_0-\alpha_2 \xi^y_0
-\dot{\alpha_1}\hat{T}_0 ,\nonumber\\
N_1 &=&  \xi^{y}_2+ \beta_1 \xi^y_1 + \beta_2 \xi^y_0 + 
\dot{\beta}_1 \hat{T}_0, \nonumber\\
B_1 &=&  e^{-2 \alpha_0}(-2 \dot{\xi}^y_0 +2 \dot{\alpha}_0 \xi^y_0+
2(\alpha_1-\beta_1)\hat{T}_0 ),
\nonumber\\
E_1 &=& \hat{E}_1 -e^{-2 \alpha_0} \xi^y_0.
\label{3-b-10}
\end{eqnarray}

\hspace{1cm}\\
(3)We take the perturbed energy momentum in the 5D spacetime as
\begin{equation}
\delta T^M_N= 
\left(
\begin{array}{ccc}
0 & 0 & 0 \\
0 & -\delta \rho & -(\rho+p) e^{\alpha_0}v_{,i} \\
0 & (\rho+p)e^{-\alpha_0} v_{,i}  & \delta p \: \delta_{ij} 
\end{array}
\right) \: \delta (y),
\label{3-b-11}
\end{equation}
where we assume the anisotropic stress of the matter is zero.
The jump of the first derivative of the metric perturbations
should be equated with the matter perturbations on the brane. 
This junction condition is obtained as (see Appendix B)
\begin{eqnarray}
\Psi_1 &=& -\alpha_1 N_0 + \frac{1}{6} \kappa^2 \delta \rho, \nonumber\\
\Phi_1 &=&  \beta_1 N_0 + \kappa^2 \left(\frac{\delta \rho}{3}+
\frac{\delta p}{2} \right), \nonumber\\
B_1 &=&  -2 (\beta_1-\alpha_1) e^{-\alpha_0} v ,\nonumber\\
E_1 &=& 0.
\label{3-b-12}
\end{eqnarray}
Combining (\ref{3-b-10}) and (\ref{3-b-12}), 
we can write the perturbed energy momentum tensor 
in terms of $\xi^y$ and $E_{m}(\k)$;
\begin{eqnarray}
\kappa^2 \delta \rho &=& -6 \left( \dot{\alpha}_0 \dot{\xi}^y_0 
- \dot{\alpha}_0^2 \xi^y_0 + \dot{\alpha}_1 \hat{T}_0\right),\nonumber\\
\kappa^2 \delta p &=& 2 \left( \ddot{\xi}_0^y + 2 \dot{\alpha}_0 
\dot{\xi}_0^y -(2 \ddot{\alpha}_0+3 \dot{\alpha}_0^2) \xi_0^y +
(\dot{\beta}_1+2 \dot{\alpha}_1)\hat{T}_0 \right), \nonumber\\
\kappa^2(\rho+p)e^{\alpha_0} v &=& 2 \dot{\xi}^y_0 -2 \dot{\alpha}_0 \xi^y_0
+\kappa^2 (\rho+p)\hat{T}_0 , 
\nonumber\\
0  &=& \hat{E}_1 -  e^{-2 \alpha_0} \xi^y_0.
\label{3-b-13}
\end{eqnarray}

\hspace{1cm}\\
(4)Imposing the constraints on the matter determines the unknown function 
$\xi^y$ and $E_{m}$. First $\xi^y_0$ is determined by the 
shareless condition;
\begin{equation}
\xi_0^y = e^{2 \alpha_0} \hat{E}_1. 
\label{3-b-14}
\end{equation}
The coefficient $E_{m}$ is determined demanding the
equation of state $\delta p = c_s^2 \delta \rho$. 
In the coordinate ($t,y)$, $\hat{E}$ can be written as
\begin{equation}
\hat{E}(t,y) = e^{-2 \alpha(y,t)}\int dm 
H^{(1)}_2(ml e^{-\alpha(y,t)}) l^2 E_m 
e^{-i m \tau(t,y)}.
\end{equation}
Let us take the limit $ml e^{-\alpha} \ll 1$. 
Then we can use the asymptotic form of the Hunkel function for small argument;
\begin{equation}
H^{(1)}_2 (z) \sim \frac{2}{z^2}+ \frac{1}{2}+{\cal O}(z^2,z^4 \cdot \cdot),
\end{equation}
where we neglect the over-all numerical coefficient.
$\hat{E}$ can be evaluated as
\begin{equation}
\hat{E}(t,y) = \int dm 
\left(\frac{2}{m^2 l^2}+\frac{1}{2}e^{-2 \alpha(y,t)}\right)
l^2 E_m 
e^{-i m \tau(t,y)},
\label{3-c-2}
\end{equation}
Using the Jacobian of the transformation (\ref{1-3}) 
\begin{eqnarray}
\left. \frac{\partial \tau}{\partial y} \right  \vert_{y=0}
&=& l(-f'(t/l)+g'(t/l))=
l \dot{\alpha}_0 e^{-\alpha_0}, \quad
\left.
\frac{\partial z}{\partial y} \right \vert_{y=0} =-l (f'(t/l)+g'(t/l))=
-l \alpha_1 e^{-\alpha_0}, \nonumber\\
\left. \frac{\partial \tau}{\partial t} \right \vert_{y=0}
&=& l(f'(t/l)+g'(t/l))=
l \alpha_1  e^{-\alpha_0},\quad \:\:\:\:
\left.\frac{\partial z}{\partial t} \right \vert_{y=0}
= l(f'(t/l)-g'(t/l))=
-l \dot{\alpha}_0  e^{-\alpha_0} , \nonumber\\
\end{eqnarray}
we can verify the following equations about 
$\xi^y_0=e^{2 \alpha_0}\hat{E}_1$
and $\hat{T}_0=-e^{2 \alpha_0}\dot{\hat{E}}_0$;
\begin{eqnarray}
\alpha_1 \xi^y_0=\beta_1 \xi^y_0 &=& 
\int dm  \left(
-1+ 2 i \frac{\dot{\alpha}_0}{m e^{-\alpha_0}}
\right) E_m e^{-i  m \tau}, \nonumber\\
\alpha_1 \dot{\xi}^y_0 &=& \int d m
\left(-2 \dot{\alpha}_0 
-i m e^{-\alpha_0} +2i \frac{\ddot{\alpha}_0+\dot{\alpha}_0^2}
{m e^{-\alpha_0}}\right)E_m
e^{-i m \tau}, \nonumber\\
\alpha_1 \ddot{\xi}^y_0 &=& \int d m
\left(-4 \ddot{\alpha}_0 -2 \dot{\alpha}_0^2 +m^2 e^{-2 \alpha_0}
- i m \dot{\alpha}_0 e^{-\alpha_0} +2i 
\frac{\dot{\ddot{\alpha}}_0+3 \dot{\alpha}_0 \ddot{\alpha}_0
+\dot{\alpha}_0^3}{m e^{-\alpha_0}}
\right)
E_m e^{-i m \tau}, \nonumber\\
\label{ee-1}
\end{eqnarray}
and 
\begin{eqnarray}
\dot{\alpha_0} \hat{T}_0 &=& \int dm \left(-2i \frac{\dot{\alpha}_0}{m e^{-\alpha_0}} \right)
E_m e^{-i m \tau} ,\nonumber\\
\dot{\hat{T}}_0 &=& \int dm 
\left(2 - 2i \frac{\dot{\alpha}_0}{m e^{-\alpha_0}} \right)
E_m e^{-i m \tau},
\label{ee-2}
\end{eqnarray}
where we used 
$\dot{\alpha_0}^2 l^2, \ddot{\alpha}_0 l^2 \sim (H l)^2  \ll 1$.
Then, we obtain
\begin{eqnarray}
\kappa^2 \alpha_1 \delta \rho &=& \int dm \left(
6 \dot{\alpha}_0^2 +6 im \dot{\alpha}_0 e^{-\alpha_0} \right) 
E_m e^{-i m \tau} ,\nonumber\\
\kappa^2 \alpha_1 \delta p &=&  \int dm 
\left(
6 w \dot{\alpha}_0^2 -6 im \dot{\alpha_0} 
e^{-\alpha_0}+ 2 m^2 e^{-2 \alpha_0} \right)
E_m e^{-i m \tau},
\end{eqnarray}
where $p=w \rho$. We assume $c_s^2=w=cosnst.$ Then we observe that
 the equation of state $\delta p=w \delta \rho$ is satisfied 
if $m e^{- \alpha_0} \ll H$.
It implies that $E_m$ should select the modes with $m e^{-\alpha_0}\ll H$.
Note that the assumption $ mle^{-\alpha_0} \ll 1$
is consistent with the result that only the modes with
$m e^{-\alpha_0}< H$ contribute to the perturbations
for late times $H < l^{-1}$. 

\subsection{Evolution of Perturbations at Superhorizon Scales}

Let us evaluate the metric perturbations $\Phi_0$ and
$\Psi_0$ in the brane world (\ref{3-b-9}).
From (\ref{3-b-9}), (\ref{ee-1}) and (\ref{ee-2}), we obtain 
\begin{equation}
\Phi_0=\Psi_0= \int dm E_{m} e^{-i m \tau}.
\label{ee-4}
\end{equation}
For late times where the Hubble horizon is sufficiently larger
than the curvature scale of the AdS spacetime $H < l^{-1}$, 
$\tau(t)$ is given by (see Appendix A)
\begin{equation}
\tau(t) \sim l e^{-\alpha_0} \left(\frac{t}{l}\right).
\end{equation}
Because $E_m$ selects the modes with $m e^{-\alpha_0} \ll H$, we find
that $m \tau \ll 1$. Then the metric fluctuations are constant.
The density fluctuations becomes 
\begin{equation}
4 \pi G_4 \delta \rho= - 3 \dot{\alpha}_0^2 \Phi_0,
\end{equation}
where we use $\kappa^4 \sigma=48 \pi G_4$.
We finally obtain the metric perturbations and density fluctuations 
on the brane  
\begin{eqnarray}
\frac{\delta \rho}{\rho}= -2 \Phi_0 = const.
\end{eqnarray}
This is one of the main result of our work.
We notice that the solutions for metric perturbations and
density perturbations are identical with those obtained in 
the conventional 4D cosmology. Note that, if $w \neq c_s^2$, 
the modes with $m e^{-\alpha_0} > H$ should contribute to 
$\Phi_0$ in order to satisfy $\delta p = c_s^2 \delta \rho$.
Once these modes contribute to $\Phi_0$, $\Phi_0$ is no longer constant. 
This also agrees with the 4D cosmology.

%%%%%%%%%%%%%%%%%%%%%%%%%%%%%%%%%%%%%%%%%%%%%%%%%%%%%%%%%%%%%%%%
%%%%%%%%%%%%%%%%%%%%%%%%%%%%%%%%%%%%%%%%%%%%%%%%%%%%%%%%%%%%%%%%
%%%%%%%%%%%%%%%%%%%%%%%%%%%%%%%%%%%%%%%%%%%%%%%%%%%%%%%%%%%%%%%%
%%%%%%%%%%%%%%%%%%%%%%%%%%%%%%%%%%%%%%%%%%%%%%%%%%%%%%%%%%%%%%%%
\section{Evolution of Perturbations at Subhorizon Scales}
\setcounter{equation}0

In this section we shall study the evolution of perturbations
at subhorizon scales $\k e^{-\alpha_0} > H$. 
We derive the equation to determine
$E_{m}(\k)$ for each mode with $\k$. 
We again use the assumption that the modes with 
$m e^{-\alpha_0}> l^{-1}$ do not contribute to the
perturbations in the brane world.
Then we investigate the late time evolution of 
perturbations at subhorizon scales larger than the
AdS curvature scale $l$ ($l^{-1}> \k e^{-\alpha_0} >H)$.
\subsection{Calculations of Perturbations}

\hspace{1cm}\\
(1) We start with the perturbed AdS spacetime
\begin{equation}
ds^2= \left(\frac{l}{z} \right)^2
\left(dz^2 - (1+2 \phi) d \tau^2 
+2 b_{,i} dx^i d \tau
+ \left((1 - 2 \hat{\Psi}) \delta_{ij}+ 
2 \hat{E}_{,ij} \right)dx^i dx^j\right), 
\end{equation}
where $\phi,b,\hat{\Psi}$ and $\hat{E}$ is given by (\ref{B-2-6}).

\hspace{1cm}\\
(2) 
The perturbations in $(y,t,x^i)$ coordinate is obtained by the coordinate
transformation. The resulting metric becomes 
(see (\ref{B-2-8}) and (\ref{B-2-9}))
\begin{eqnarray}
ds^2 &=&  e^{2 \beta(y,t)} \left((1+2 \hat{N}) dy^2 -(1+2 \hat{\Phi})
d t^2 +2 \hat{A} \: dt \: dy \right) \nonumber\\
&& +  e^{2 \alpha(y,t)}
\left(\left((1 - 2 \hat{\Psi}) \delta_{ij}+ 2 \hat{E}_{,ij}\right) 
dx^i dx^j+2 \hat{B}_{,i} dx^i dt + 2 \hat{G}_{,i} dx^i dy \right),
\end{eqnarray}
where
\begin{eqnarray}
\hat{\Phi} &=&  (l \alpha')^2 e^{-2 \beta} \phi, \nonumber\\
\hat{B} &=&  (l \alpha') e^{-\alpha} b ,\nonumber\\
\hat{N} &=& - (l \dot{\alpha})^2 e^{-2 \beta} \phi ,\nonumber\\
\hat{A} &=& -2 (l^2 \dot{\alpha} \alpha') e^{-2 \beta} \phi ,\nonumber\\
\hat{G} &=& (l \dot{\alpha}) e^{- \alpha} b.
\end{eqnarray}
In addition, we perform the (infinitesimal) coordinate transformation by
\begin{equation}
x^M \to x^M + \xi^M, \quad \xi^M=(\xi^y,\xi^t,\xi^{,i}).
\end{equation}
We will take the gauge condition $G=A=0$ and $B_0=0,E_0=0$. 
This determines $\xi^t$ and $\xi$ in terms of $\xi^y$ as
\begin{eqnarray}
\xi^t &=& \int^y_0 dy (\hat{A}+ \dot{\xi}^y) + \hat{T}_0, \quad \quad 
\hat{T}_0 = e^{2 \alpha_0} (\hat{B}_0-\dot{\hat{E}}_0), \nonumber\\
\xi &=& -\int^y_0 dy (\hat{G}+e^{2(\beta-\alpha)} \xi^y) 
-\hat{E}_0.
\end{eqnarray}
Then we obtain the metric perturbations on the brane
\begin{eqnarray}
\Phi_0 &=& \hat{\Phi}_0+ 
\beta_1 \xi^y_0 +\dot{\hat{T}}_0, \nonumber\\
\Psi_0 &=& \hat{\Psi}_0- \alpha_1 \xi_0^y -\dot{\alpha}_0 \hat{T}_0, \nonumber\\
N_0 &=& \hat{N}_0 + \xi^{y}_1 + \beta_1 \xi^y_0, 
\label{5-1}
\end{eqnarray}
and the first derivative of the metric perturbations
\begin{eqnarray}
\Phi_1 &=& \ddot{\xi}^y_0 + \beta_1 \xi^{y}_1+ \beta_2 \xi_0^y 
+ \hat{\Phi}_1 + \dot{\hat{A}}_0+ \dot{\beta}_1 \hat{T}_0,
\nonumber\\
\Psi_1 &=& - \alpha_1 \xi^{y}_1-\dot{\alpha}_0 \dot{\xi}^y_0-\alpha_2 \xi^y_0
+ \hat{\Psi}_1 - \dot{\alpha}_0 \hat{A}_0 -\dot{\alpha}_1 \hat{T}_0
,\nonumber\\
N_1 &=&  \xi^{y}_2+ \beta_1 \xi^y_1 + \beta_2 \xi^y_0 +
\hat{N_1}+ \dot{\beta}_1 \hat{T}_0 
,\nonumber\\
B_1 &=&  e^{-2 \alpha_0}(-2 \dot{\xi}^y_0 +2 \dot{\alpha}_0 \xi^y_0
- 2 (\beta_1-\alpha_1) \hat{T}_0 
- \hat{A}_0 +e^{2 \alpha_0} \hat{B}_1 - e^{2 \alpha_0} \dot{\hat{G}}_0
) ,\nonumber\\
E_1 &=& \hat{E}_1 -e^{-2 \alpha_0} \xi^y_0 -\hat{G}_0.
\label{5-2}
\end{eqnarray}

\hspace{1cm}\\
(3)Combining the junction conditions (\ref{3-b-12}) and (\ref{5-2}),
we can write matter in terms of
$\xi^y$ and $E_{m}(\k)$;
\begin{eqnarray}
\kappa^2 \delta \rho &=& -6 \left( \dot{\alpha}_0 \dot{\xi}^y_0 
+(\alpha_2 -\alpha_1 \beta_1) \xi^y_0 
-\hat{\Psi}_1 + \dot{\alpha}_0 \hat{A}_0 + \dot{\alpha}_1 \hat{T}_0
-\alpha_1 \hat{N}_0 \right), \nonumber\\
\kappa^2 \delta p &=& 2 \left( \ddot{\xi}_0^y + 2 \dot{\alpha}_0 
\dot{\xi}_0^y 
+(2 \alpha_2 + \beta_2 -\beta_1^2 -2 \alpha_1 \beta_1)\xi_0^y 
\right. \nonumber\\
&& \left. + \hat{\Phi}_1 -2 \hat{\Psi}_1 + \dot{\hat{A}}_0
+2 \dot{\alpha}_0 \hat{A}_0 
+(\dot{\beta}_1+2 \dot{\alpha}_1) \hat{T}_0 
- (\beta_1+2 \alpha_1)\hat{N}_0 \right), \nonumber\\
\kappa^2(\rho+p) e^{\alpha_0} v &=& 2 \dot{\xi}_0^y -2 \dot{\alpha}_0 \xi^y_0 
- e^{2 \alpha_0} \hat{B}_1 
+ 2 (\beta_1-\alpha_1) \hat{T}_0 +e^{2 \alpha_0} \dot{\hat{G}}_0 + \hat{A}_0 
,\nonumber\\
0&=& -2 e^{-2 \alpha_0} \xi^y_0 +2 \hat{E}_1 -2 \hat{G}_0.
\label{5-3}
\end{eqnarray}

\hspace{1cm}\\
(4)Imposing the condition on 
the anisotoropic stress $\pi_T$ 
and the equation of state $c_s^2= \delta p/\delta \rho$
determines $\xi^y_0$ and $E_{m}(\k)$. Substituting
$\xi^y_0=e^{2 \alpha_0}(\hat{E}_1 - \hat{G}_0)$ 
into $\delta p -c_s^2 \delta \rho=0$, we obtain the following form
of the equation for each $\k$;
\begin{equation}
\int dm F(t, m ; \k) E_{m}(\k) =0.
\end{equation}
Because the detailed form of $F(t, m ; \k)$ is rather complicated,
we omit it here. 
Defining the Fourier transformation of the function $F(t, m ;\k)$ by
\begin{equation}
F(t, m ;\k) = \int d m' \tilde{F}(m',m ; \k)
e^{-i m t},
\end{equation}
this equation becomes
\begin{equation}
\int dm \tilde{F}(m', m; \k) E_{m}(\k) =0.
\label{ee-3}
\end{equation}
The problem is to find the eigenstate of the matrix $F(m',m)$
with eigenvalue $0$. Since $F(t, m ;\k)$ is consisted from the
combination of the oscillating function, the existence of the
solution is very likely.

\subsection{Evolution of Perturbations at Subhorizn Scales}
The equation to determine $E_{m}(\k)$ is rather complicated. 
However we can deduce the evolution of perturbations 
for late times $H < l^{-1}$ from the following arguments. 
We take the assumption that the massive modes with 
$m e^{-\alpha_0} > l^{-1} $ can be neglected. 
We can evaluate the metric perturbations as
\begin{eqnarray}
\Phi_0 &=& \int dm \frac{3 m^2}{2 \k^2 +3 m^2}
\left(1 + \frac{1}{6} (\k l e^{-\alpha_0})^2 \right) E_{m}(\k)
e^{-i \omega \tau},\nonumber\\
\Psi_0 &=& \int dm \frac{3 m^2}{2 \k^2 +3 m^2}
\left(1 - \frac{1}{6} (\k l e^{-\alpha_0})^2 \right) E_{m}(\k)
e^{-i \omega \tau},
\end{eqnarray}
where we neglect the terms of the order 
${\cal O}((m l e^{-\alpha_0})^2)$.
Then we find that
\begin{equation}
\Phi_0-\Psi_0 = {\cal O}((\k l e^{-\alpha_0})^2).
\end{equation}
For $\k l e^{-\alpha_0} \to 0$ but not necessarily 
$ \k e^{-\alpha_0} <H $, these agree with the result 
obtained in (\ref{ee-4}). 
A notable point is that this equation is sufficient to 
show that the 4D cosmology is reproduced.
From the 5D Einstein equation, we have alredy obtained 
the three equations for $\Phi_0$, $\Psi_0$, $\delta \rho$ and 
$v$ (conservation of the energy momentum (\ref{ee-5})
and trace-part of the Einstein equation (\ref{ee-6})).
Thus, for $\k l e^{-\alpha_0} \ll 1$, we have closed set of equations about 
the metric perturbations and density fluctuations, which is identical 
with the one obtained in the conventional 4D equation. Then we conclude that
if the effect of the 
massive modes with $m l e^{-\alpha_0} > 1$ can be neglected,
the cosmological perturbations are not interfered by the extra dimension.
Whether the modes with $m l e^{-\alpha_0} > 1$ contribute to the 
perturbations or not is determined by $E_{m}(\k)$ which can be obtained 
from (\ref{ee-3}).

%%%%%%%%%%%%%%%%%%%%%%%%%%%%%%%%%%%%%%%%%%%%%%%%%%%%%%%%%%%%%%%%
%%%%%%%%%%%%%%%%%%%%%%%%%%%%%%%%%%%%%%%%%%%%%%%%%%%%%%%%%%%%%%%%
%%%%%%%%%%%%%%%%%%%%%%%%%%%%%%%%%%%%%%%%%%%%%%%%%%%%%%%%%%%%%%%%
%%%%%%%%%%%%%%%%%%%%%%%%%%%%%%%%%%%%%%%%%%%%%%%%%%%%%%%%%%%%%%%%
\section{Deviation from 4D cosmology.}
\setcounter{equation}0
In the previous two sections, we show that the late time
evolution of the perturbations is not modified by the
5D graviton if the the effect of the modes with
$m e^{-\alpha_0} \gg l^{-1}$ is negligible.
Let us investigate the effect of these massive modes.
To simplify the calculation, we assume $\k \to 0$. 
Without taking the limit $m l e^{- \alpha_0} \ll 1$, 
$\alpha_1 \xi^y_0$ and $\dot{\alpha}_0 \hat{T}$ becomes
\begin{eqnarray}
\alpha_1 \xi^y_0 &=& - \int dm (m l e^{-\alpha_0})E_m 
H^{(1)}_1(m l e^{-\alpha_0})e^{-im \tau} \nonumber\\
&&+ \dot{\alpha}_0 e^{-\alpha_0} \int dm (im l^2) E_m 
H^{(1)}_2(m l e^{-\alpha_0})e^{-im \tau}
,\nonumber\\
\dot{\alpha}_0 \hat{T}_0 &=& \dot{\alpha}_0^2 l^2 
\int dm (m l e^{-\alpha_0}) E_m H^{(1)}_1(m l e^{-\alpha_0})e^{-im \tau} \nonumber\\
&&- \dot{\alpha}_0 e^{-\alpha_0} \int dm (im l^2) E_m 
H^{(1)}_2(m l e^{-\alpha_0})e^{-im \tau}. 
\end{eqnarray}
Then we can evaluate $\Psi_0$ as
\begin{equation}
\Psi_0 = (1- \dot{\alpha_0}^2 l^2) 
\int dm \Psi_{m}(t)
E_{m}(\k) e^{-i \omega \tau}, \quad
\Psi_{m}(t) = (m l e^{-\alpha_0}) H_1^{(1)}(m l e^{-\alpha_0}).
\label{F-1}
\end{equation}
We shall investigate the effect of the massive modes.
The evolution of $\Psi_{m}$ can be estimated using
the asymptotic form of the Hunkel function 
for large argument $H^{(1)}_1(z) \propto 1/\sqrt{z}$. 
We find that $\Psi_{m}(t)$ behaves as
\begin{equation}
\Psi_{m}(t) \propto e^{-\alpha_0(t)/2},\quad (m > l^{-1} e^{\alpha_0}).
\end{equation}
Then, the modes with
\begin{equation}
m > m_{eff}= l^{-1}_{eff}, \quad l_{eff}= l e^{-\alpha_0}, 
\end{equation}
modify the evolution of the metric perturbations and 
hence of the density fluctuations.
Because we have normalized the scale factor as 
$e^{\alpha_0(t=t_{present})}=1$, $l_{eff}$ becomes larger than 
$l$ for early times. 

The result can be interpreted as follows. For late times,
the trajectory of the brane is identical with the RS brane (see Fig.2).
Thus the behavior of the gravity in the brane world for late times 
can be deduced from the
arguments for RS solution. In the coordinate system $(z,\tau,x^i)$, the
perturbations $h(z,\tau, x^i)$ satisfies the wave equation (\ref{B-2-3}). 
Defining 
$h(z,\tau, x^i)= (z/l)^{3/2} \psi(z)e^{-i \omega \tau} e^{i \k \x}$,
the wave equation can be rewritten as
\begin{equation}
\left(- \frac{1}{2}\frac{d^2}{d z^2} + V_{AdS}(z) \right) \psi(z)
=\frac{1}{2}m^2 \psi(z), \quad 
V_{AdS}(z)= \frac{15}{8} \frac{1}{z^2}.
\end{equation}
The problem can be understood as the potential problem in 
one dimensional quantum mechanics where $m$ represents the energy.
If we cut the AdS spacetime at $z=l$ and 
put the brane there, the potential term proportional to $\delta(z-l)$
appears. Imposing the $Z_2$ symmetry across the brane, 
the potential becomes a ``volcano`` potential. 
The term proportional to the delta function is responsible for the
normalizable zero-mode of the 5D graviton which reproduces 
the 4D gravity on the brane. 
The brane is protected from the 
the massive modes by the potential barrier $V_{AdS}$ (see Fig.4).
Only the modes with large $m$ can affect the gravity on the brane. 
Let us consider the cosmological brane for early times. 
For early times, the brane is located at large $z>l$. For larger $z$, 
the potential barrier $V_{AdS}$ is lower, then the mode with 
smaller $m$ can affect the gravity on the brane. This picture 
is consistent with the result that the modes with large $m > m_{eff}
=l^{-1} e^{\alpha_0}$ can modify the evolution of the metric 
perturbations because $m_{eff}$ becomes smaller for early times.

\begin{figure}[t]
  \epsfysize=80mm
\begin{center}
  \epsfbox{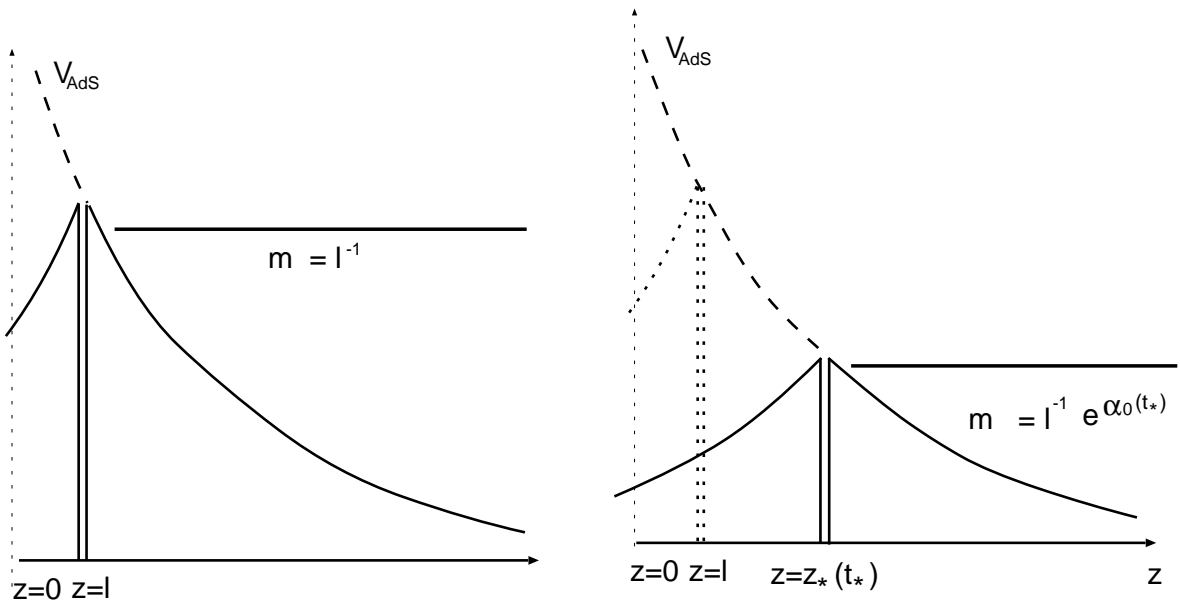}
\end{center}
  \caption{Volcano potential for 5D gravity. The location of the brane 
determines the height of the ''barrier'' which protects the 
brane from massive modes.}  
  \label{fig3}
\end{figure}

%%%%%%%%%%%%%%%%%%%%%%%%%%%%%%%%%%%%%%%%%%%%%%%%%%%%%%%%%%%%%%%%%
%%%%%%%%%%%%%%%%%%%%%%%%%%%%%%%%%%%%%%%%%%%%%%%%%%%%%%%%%%%%%%%%%
%%%%%%%%%%%%%%%%%%%%%%%%%%%%%%%%%%%%%%%%%%%%%%%%%%%%%%%%%%%%%%%%%
%%%%%%%%%%%%%%%%%%%%%%%%%%%%%%%%%%%%%%%%%%%%%%%%%%%%%%%%%%%%%%%%%

\section{Discussions}                  
\setcounter{equation}0
The cosmological perturbations in the brane world provide 
useful tests for the brane world idea. This is because the perturbations
in the brane world interact with the perturbations in the bulk
which is inherent nature of perturbations in the brane world.
The dynamics of the brane can not be separated from the dynamics of the bulk.  
This is because the inhomogeneous
fluctuations on the brane inevitably produces the gravitational
waves in the bulk, which in turn affect the evolution of 
perturbations on the brane. Thus, naively, we think the evolution of the 
cosmological perturbations is modified significantly. 

We showed that this is not a case for late time $H < l^{-1}$ and
at scales larger than the AdS curvature scale.
The metric perturbations become frozen once the
perturbations exit the horizon as in the conventionoal 4D cosmology.
This result is important for the inflationary scenario in the brane
world. If the scale $l^{-1}$ is sufficiently higher than the scales
of the inflation, and if heavy graviton modes may be neglected 
the constancy of the curvature perturbations can be used to estimate 
the scalar temperature anisotropies of the CMB at large scales. 
Our results are consistent with those
of Ref \cite{MWBH}, where curvature perturbations on large scales is
shown to be conserved, and the density perturbations generated during 
high-energy inflation on the brane are calculated.

The assumption in obtaining the above results is that 
the effect of massive gravitons with 
$m >m_{eff}= e^{\alpha_0}l^{-1}$ can be neglected.
The contribution of these modes depends on the initial
spectrum of the fluctuations. If the primordial fluctuations 
are generated during inflation at low energies $H <l^{-1}$, 
heavy gravitions are significantly suppressed \cite{pre}.
Then, in this case, the assumption seems to be natural. 

Another modification of the evolution 
arises if the scales of the perturbations becomes 
smaller than $l$. This
modification is not important for late time evolution because the
cosmological scales is significantly larger than $l$. However at the
beginning of the inflation, we should consider the scales comparable to
$l$. Then the modification becomes important to predict the primordial
spectrum of the fluctuations during inflation. 

The key to quantify 
these modifications is the understanding of $E_{m}(\k)$.
The equation for $E_{m}(\k)$ obtained in this paper
might give a way to estimate the effect of the interaction 
with bulk graviton on the perturbations in the brane world, 
which is the intrinsic feature of the brane world cosmology. 

Finally, we comment on the possibility for the extension of the 
present work. In this paper, we have considered the spatially flat
universe. The extension to the non-flat universe may be possible using the  
coordinate system in Ref \cite{BWC10}. The extension of
our method to vector and tensor perturbations is straightforward. 
The tensor perturbations were calculated in the de Sitter 
background and the agreement with 4D gravity was demonstrated
in a certain limit \cite{HHR}. It will be interesting 
to investigate the effect of the vector components of 5D graviton 
on the 4D cosmological perturbations.  

\hspace{1cm}\\

While the present work was being completed, related papers 
Ref [26-29] appear on the hep-th. 

\section*{Acknowledgements}
We would like to thank S. Kawai, T. Shiromizu and T. Tanaka 
for helpful comments.
The work of K.K. was supported by JSPS Research Fellowships for
Young Scientist No.4687.

%%%%%%%%%%%%%%%%%%%%%%%%%%%%%%%%%%%%%%%%%%%%%%%%%%%%%%%%%%%%%%%
%%%%%%%%%%%%%%%%%%%%%%%%%%%%%%%%%%%%%%%%%%%%%%%%%%%%%%%%%%%%%%%
%%%%%%%%%%%%%%%%%%%%%%%%%%%%%%%%%%%%%%%%%%%%%%%%%%%%%%%%%%%%%%%
%%%%%%%%%%%%%%%%%%%%%%%%%%%%%%%%%%%%%%%%%%%%%%%%%%%%%%%%%%%%%%%

\appendix

\section{Background - Einstein equation and junction condition}
\setcounter{equation}0
In this appendix, we derive the junction conditions and
obtain a solution for the background.
The junction conditions can be obtained from the 5D Einstein equation;
\begin{equation}
G^M_N= \frac{6}{l^2} \delta^M_N + \kappa^2 
\frac{\sqrt{-g_{brane}}}{\sqrt{-g}} T^M_N 
=  \frac{6}{l^2} \delta^M_N 
+ e^{-\beta} \kappa^2 T^M_N  , 
\:\: (M,N =y,t,x^i). 
\label{1-7}
\end{equation}
We take for the energy momentum tensor in the 5D spacetime 
\begin{equation}
T^M_N= \left(- \sigma diag(0,1,1,1,1)+
diag(0,-\rho,p,p,p) \right) \delta(y). 
\end{equation}
The jump of the first derivative of $\alpha(y,t)$ and 
$\beta(y,t)$ gives the $\delta(y)$ function to the Einstein tensor.
The Einstein tensor is given by
\begin{eqnarray}
G^0_{\: 0} &=& - 3 e^{-2 \beta}(\dot{\alpha}^2 + \dot{\alpha} \dot{\beta} 
-\alpha''-2 \alpha'^2 + \alpha' \beta') ,\nonumber\\
G^y_{\: y} &=& 3 e^{-2 \beta}
( -\ddot{\alpha}-2 \dot{\alpha}^2 + \dot{\alpha} \dot{\beta}
+ \alpha'^2 + \alpha' \beta') ,\nonumber\\
G^0_{\: y} &=&  -3 e^{-2 \beta}
( \beta' \dot{\alpha}+ \alpha' \dot{\beta} - \dot{\alpha}'
-\dot{\alpha} \alpha') ,\nonumber\\
G^{i}_{\:j} &=& 
\delta^{i}_{j} e^{-2 \beta}(-2 \ddot{\alpha} -3 \dot{\alpha}^2
- \ddot{\beta} + 2 \alpha''+ 3 \alpha'^2 + \beta'') .
\end{eqnarray}
Equating the coefficient of $\delta(y)$ we obtain
the junction conditions
\begin{eqnarray}
\alpha_1(t) &=& - \kappa^2 e^{\beta_0} \left(\frac{\sigma}{6}   
+\frac{\rho(t)}{6} \right), \nonumber\\
\beta_1(t) &=&  - \kappa^2 e^{\beta_0} 
\left(\frac{\sigma}{6}-\frac{\rho(t)}{3}-\frac{p(t)}{2} \right).
\end{eqnarray}
Following Ref \cite{BWC5}, we make power series expansion
of the Einstein equation near the brane. 
The $y^0$ order of the $(y,y)$ ,$ (y,0)$ 
component of the Einstein equation gives 
\begin{eqnarray}
&& \ddot{\alpha}_0+2 \dot{\alpha}_0^2= 4 \pi G_4 
\left(\frac{\rho}{3}-p \right) - \frac{\kappa^4 \rho(\rho+3 p)}{36},
\nonumber\\
&& \dot{\rho}+ 3 \dot{\alpha}_0 (\rho+p)=0,
\end{eqnarray}
where $\kappa^4 \sigma=48 \pi G_4$. 
The integration of the first equation gives 
\begin{equation}
\dot{\alpha}_0^2 = \frac{8 \pi G_4}{3} \rho + 
\frac{\kappa^4 \rho^2}{36} + e^{-4 \alpha_0} C ,
\end{equation}
where $C$ is the constant of the integration and is proportional to the
mass of the AdS-Schwarzshild mass. Because the bulk is AdS spaceatime
in our solution, $C=0$. These equations are equivalent with (\ref{1-16}).
The $y^0$ order of $(0,0)$ and $(i,j)$ components of the Einstein equation
give $\alpha_2$ and $\beta_2$ in terms of $\alpha_0$, $\alpha_1$,
$\beta_0$ and $\beta_1$;
\begin{eqnarray}
\alpha_2 &=& \dot{\alpha}_0^2 + \dot{\alpha}_0 \dot{\beta}_0 
-2 \alpha_1^2 + \alpha_1 \beta_1 + \frac{2}{l^2},\nonumber\\
\beta_2  &=& \dot{\alpha}_0^2+2 \ddot{\alpha}_0+ \ddot{\beta}_0+\alpha_1^2
-2 \alpha_1 \beta_1 + \frac{2}{l^2}.
\end{eqnarray}

Let us find the function $f(u)$ and $g(v)$ for late times. 
For late times, we can neglect ${\cal O}((\rho/\sigma)^2)$. 
Then the solution for $e^{-\alpha_0}$ is given by
\begin{equation}
e^{-\alpha_0}=f(t/l)-g(t/l)= a_{\ast} \left(\frac{t}{l} \right)^{-\frac{2}{3(1+w)}}.
\end{equation}
Combining this equation with $e^{2 \beta_0}=1$;
\begin{equation}
 4 \frac{f'(t/l)g'(t/l)}{(f(t/l)-g(t/l))^2}=1,
\end{equation}
we can obtain the first order differential equation
for $f(t/l)$;
\begin{equation}
\dot{f}(t/l)=\frac{a_{\ast}}{3(1+w)}
\left(\frac{t}{l}\right)^{-\frac{5+3w}{3(1+w)}} 
\left(-1 + \sqrt{1+ \left(\frac{3}{2}(1+w) \frac{t}{l} \right)^2} \right).
\end{equation}
For late times $t/l \gg1$, we obtain the solution for $f(t)$ and $g(t)$;
\begin{eqnarray}
f(t/l) &=& a_{\ast} \frac{1}{2} \left(\frac{t}{l}\right)^{- 
\frac{2}{3(1+w)}}\left(
1+ \frac{3(1+w)}{1+3 w}\frac{t}{l}
 \right)+b_{\ast}, \nonumber\\
g(t/l) &=& a_{\ast} \frac{1}{2} \left(\frac{t}{l}\right)^{- 
\frac{2}{3(1+w)}} \left(
-1+ \frac{3(1+w)}{1+3 w}\frac{t}{l}
 \right)+b_{\ast}, 
\end{eqnarray}
where $a_{\ast}$ and $b_{\ast}$ is the constants of integration.
Then we obtain 
\begin{equation}
\tau(t) \sim l e^{-\alpha_0} \left(\frac{t}{l}\right),
\end{equation}
where we take $b_{\ast}=0$.

%%%%%%%%%%%%%%%%%%%%%%%%%%%%%%%%%%%%%%%%%%%%%%%%%%%%%%%%%%%
%%%%%%%%%%%%%%%%%%%%%%%%%%%%%%%%%%%%%%%%%%%%%%%%%%%%%%%%%%%
%%%%%%%%%%%%%%%%%%%%%%%%%%%%%%%%%%%%%%%%%%%%%%%%%%%%%%%%%%%
\section{Perturbations - Einstein equation and junction conditions}
\setcounter{equation}0
\subsection{Gauge fixing and Einstein tensor}

In this section, we derive the 5D Einstein equation for perturbed
AdS spacetime and derive the junctions conditions.
We will concentrate on the scalar perturbations.
We put the perturbed 5D AdS spacetime as
\begin{eqnarray}
ds^2 &=& e^{2 \beta(y,t)} \left((1+2 N) dy^2 -(1+2 \Phi)
d t^2 +2 A \: dt \: dy \right) \nonumber\\
&+&
e^{2 \alpha(y,t)}
\left(\left((1 - 2 \Psi) \delta_{ij}+ 2 E_{,ij}\right) 
dx^i dx^j+ 2 B_{,i} dx^i d t  + 
2 G_{,i} dx^i dy \right). 
\end{eqnarray}
There are three degrees of freedom in the gauge transformations 
\begin{equation}
x^M \to x^M + \xi^M, \quad \xi^M=(\xi^y,\xi^t,\xi^{,i}).
\end{equation}
By this gauge transformations, metric perturbations are transformed as
\begin{eqnarray}
\Phi &=& \hat{\Phi}+\dot{\xi}^t+ \beta' \xi^y + \dot{\beta} \xi^t 
,\nonumber\\
\Psi &=& \hat{\Psi}- \dot{\alpha} \xi^t -\alpha' \xi^y ,\nonumber\\
E &=& \hat{E} + \xi ,\nonumber\\
B &=& \hat{B} + \dot{\xi} - e^{2 (\beta-\alpha)} \xi^t, \nonumber\\
A &=& \hat{A} + \dot{\xi}^y -\xi^{t'} ,\nonumber\\
G &=& \hat{G} + e^{2(\beta-\alpha)} \xi^y+\xi' ,\nonumber\\
N &=& \hat{N} + \xi^{y'}+\dot{\beta} \xi^t + \beta' \xi^y.
\end{eqnarray}
Using these degree of the freedom, 
we impose the gauge conditions so that the resulting coordinate
becomes Gaussian Normal (GN) coordinate because the metric
perturbations in the GN coordinate on the brane is the metric
perturbations observed by the observers confined to the brane.
In the GN coordinate, the transverse component of the metric 
$g_{y \mu},(\mu=t,x^i)$ vanish $(G(y,t,x^i)=A(y,t,x^i)=0)$ 
and the brane is located at $y=0$.
The former conditions are achieved by $\xi$ and $\xi^t$
and the latter condition is achieved by $\xi^y$. 
The conditions $G=A=0$ determines $\xi^t$ and $\xi$ as
\begin{eqnarray}
\xi^t &=& \int^y_0 dy \left(\hat{A}+ \dot{\xi}^y \right)
 + \epsilon^t(t,x^i), \nonumber\\
\xi &=& -\int^y_0 dy 
\left(\hat{G}+ e^{2(\beta-\alpha)} \xi^y \right)+ \epsilon(t,x^i),
\end{eqnarray}
where $\epsilon^t$ and $\epsilon$ are functions with no $y$ dependence.
These residual gauge transformations 
enable us to impose two additional gauge fixing conditions. 
We take $B_0(y=0,t,x^i)=E_0(y=0,t,x^i)=0$ gauge on the analogy of
the longitudinal gauge in the conventional 4D cosmological perturbations
theory. The Einstein tensor is calculated as 
\begin{eqnarray}
\delta G^0_0 &=& e^{-2 \beta} \left(
(6 \dot{\alpha}^2 + 6 \dot{\alpha} \dot{\beta}) \Phi - 3 \alpha' N'
-3 \dot{\alpha} \dot{N} -( 6 \alpha''+12 \alpha'^2 - 6 \alpha' \beta')N
\right.
\nonumber\\
&& \left. - 3 \Psi'' + 3( \dot{\beta}+ 2 \dot{\alpha}) \dot{\Psi} + 
3( \beta'-4 \alpha') \Psi' \right) \nonumber\\
&& + \nabla^2 \left( e^{-2 \beta} (E'' -(\dot{\beta}+ 2 \dot{\alpha})
\dot{E}-(\beta'-4 \alpha')E') -2 e^{-2 \alpha} \Psi 
+ e^{- 2 \alpha} N \right. \nonumber\\
&& \left. + e^{-2 (\alpha+\beta)}
(\dot{\beta}+ 2 \dot{\alpha})\tilde{B}
\right), \nonumber\\
\delta G^y_{\: y} &=& e^{-2 \beta} \left( 
-(6 \alpha'^2 + 6 \alpha' \beta') N + 3 \alpha' \Phi'
+ 3 \dot{\alpha} \dot{\Phi} + ( 6 \ddot{\alpha}+12 \dot{\alpha}^2 
- 6 \dot{\alpha} \dot{\beta}) \Phi
\right.\nonumber\\
&&\left. + 3 \ddot{\Psi} - 3( \dot{\beta}- 4 \dot{\alpha}) \dot{\Psi} -
3( \beta'+ 2 \alpha') \Psi' \right) \nonumber\\
&& + \nabla^2 \left( e^{-2 \beta} (- \ddot{E} +(\dot{\beta}-4 \dot{\alpha})
\dot{E}+(\beta'+ 2 \alpha')E') -2 e^{-2 \alpha} \Psi 
+ e^{- 2 \alpha} \Phi \right. \nonumber\\
&& \left. + e^{-2 (\alpha+\beta)}(\dot{\tilde{B}}+
(2 \dot{\alpha}-\dot{\beta}))\tilde{B}
\right) ,\nonumber\\
\delta G^y_{\: 0} &=& e^{-2 \beta} \left( 3 \dot{\Psi}'+ 3 \dot{\alpha} \Phi'
+ 3 \alpha' \dot{N} -3(\beta'-\alpha') \dot{\Psi}
-3 (\dot{\beta}-\dot{\alpha}) \Psi' \right) \nonumber\\
&& + e^{-2 \beta}\nabla^2 \left( -\dot{E}'+(\beta'-\alpha') \dot{E} 
+(\dot{\beta}- \dot{\alpha})E'+ e^{-2\alpha}(- \beta' \tilde{B} +
\frac{1}{2} \tilde{B}') \right) ,\nonumber\\
\delta G^y_{\: i} &=& e^{-2 \beta} \left(
(2 \alpha'+ \beta')N + (\alpha'-\beta') \Phi - \Phi'+ 2 \Psi' 
\right. \nonumber\\ 
&&
\left.+ 
e^{-2 \beta}\left (( 3 \dot{\alpha} \alpha' + \dot{\alpha}'-2 \alpha'
\dot{\beta})\tilde{B} +( \dot{\beta}- \frac{3}{2} \dot{\alpha} )
\tilde{B}'+\alpha' \dot{\tilde{B}}
-\frac{1}{2} \dot{\tilde{B}}' \right ) \right)_{,i} ,\nonumber\\
\delta G^0_{\: i} &=& e^{-2 \beta} \left(
-(2 \dot{\alpha}+ \dot{\beta})\Phi - (\dot{\alpha}-\dot{\beta})N 
+ \dot{N} -  2 \dot{\Psi} 
\right. \nonumber\\
&& \left. + e^{-2 \beta}\left 
(-( 3 \alpha'^2 + \alpha''-2 \alpha' \beta')\tilde{B} 
-(\beta'- \frac{1}{2} \alpha' )\tilde{B}'
+\frac{1}{2} \tilde{B}'' \right ) \right)_{,i}, \nonumber\\
\delta G^i_{\: j}&=& e^{- 2 \beta} \delta^{i}_{\: j}
\left(\Phi''+ (4 \ddot{\alpha}+2 \ddot{\beta}+ 6 \dot{\alpha}^2) \Phi
+(\dot{\beta}+2 \dot{\alpha}) \dot{\Phi}+(\beta'+2 \alpha')\Phi'
-\ddot{N} \right.
\nonumber\\
&&  \left.
-( 4 \alpha''+2 \beta''+ 6 \alpha'^2)N 
 -(\beta'+2 \alpha')N'
-(\dot{\beta}+2 \dot{\alpha})\dot{N}+ 2 \ddot{\Psi} + 6 \dot{\alpha}\dot{\Psi}
- 6 \alpha' \Psi' -2 \Psi''\right) \nonumber\\
&&+ \nabla^2 \left( e^{-2 \alpha}(\Phi+ N-\Psi)+ e^{- 2 \beta} 
\left(e^{-2 \alpha}(
\dot{\tilde{B}}+\dot{\alpha}\tilde{B}) - 3 \dot{\alpha} \dot{E}
+ 3 \alpha' E' - \ddot{E}+ E'' \right) \right) \delta^i_{\: j}\nonumber\\
&& -\left( e^{-2 \alpha}(\Phi+ N-\Psi)+ e^{- 2 \beta} 
\left(e^{-2 \alpha}(
\dot{\tilde{B}}+\dot{\alpha}\tilde{B}) - 3 \dot{\alpha} \dot{E}
+ 3 \alpha' E' - \ddot{E}+ E'' \right) \right)^{,i}_{\: ,j},
\end{eqnarray}
where $\tilde{B}=e^{2 \alpha} B$ and 
we denote $h'= \partial h/\partial y$ and 
$\dot{h}= \partial h/\partial t$. 

\subsection{Einstein equation and junction conditions}
In this subsection, we derive the junction conditions. 
The Einstein equation is given by 
\begin{equation}
\delta G^M_N = 
\kappa^2 e^{- \beta} (\delta T^M_N - N T^{M}_{N}).
\end{equation}
We take for the 5D energy-momentum tensor
\begin{equation}
\delta T^M_N = 
\left(
\begin{array}{ccc}
0 & 0 & 0 \\
0 & -\delta \rho & -(\rho+p)e^{\alpha_0} v_{,i} \\
0 & (\rho+p)e^{-\alpha_0} v_{,i}  & \delta p \: \delta_{ij}\\
\end{array}
\right) \: \delta (y).
\end{equation}
By equating the coefficients of $\delta (y)$ in the Einstein
equation gives the following junction conditions
\begin{eqnarray}
\Psi_1 &=& -\alpha_1 N_0 +\frac{1}{6} \kappa^2 
\delta \rho, \nonumber\\
\Phi_1 &=&  \beta_1 N_0 + \kappa^2 
\left(\frac{\delta \rho}{3}+
\frac{\delta p}{2} \right), \nonumber\\
B_1 &=&  -2 (\beta_1-\alpha_1) e^{-\alpha_0} v, \nonumber\\
E_1 &=&  0,
\end{eqnarray}
where we use $e^{\beta_0}=1$.


\begin{thebibliography}{99}
\bibitem{BW1}
V. A. Rubakov and M. E. Shaposhnikov, Phys. Lett. {\bf B125}, 
136 (1983).

\bibitem{BW2}
K. Akama, "Pregeometry" in Lecture Notes in Physics, 176, Gauge Theory
and Gravitation, Proceedings, Nara, 1982, (Springer-Verlag), edited by K.
Kikkawa, N. Nakanishi and H. Nariai, 267-271, hep-th/0001113. 

\bibitem{RS}
L. Randall and  R. Sundrum,  Phys. Rev. Lett. {\bf 83}, 4690 (1999), 
see also L. Randall and R. Sundrum,  Phys. Rev. Lett. {\bf 83}, 3370 (1999).

\bibitem{BWC0}
H. A. Chamblin and  H. S. Reall, 
Nucl. Phys. {\bf B562}, 133 (1999).

\bibitem{BWC1}
N. Kaloper,  Phys. Rev. {\bf D60}, 123506 (1999). 

\bibitem{BWC2}
T. Nihei, Phys. Lett. {\bf B465}, 81 (1999).

\bibitem{BWC3}
H.B. Kim, H.D. Kim,  Phys. Rev. {\bf D61}, 064003 (2000).

\bibitem{BWC4}
P. Bin{\'e}truy, C. Deffayet, U. Ellwanger, 
D. Langlois, Phys. Lett. {\bf B477}, 285 (2000).

\bibitem{BWC5}
E. Flanagan, S. Tye, and I. Wasserman, 
Phys. Rev. {\bf D62}, 024011 (2000).

\bibitem{BWC6}
P. Kraus,  JHEP {\bf 9912}, (1999) 011.

\bibitem{BWC7}
D. Vollick, Class. Quant. Grav. {\bf 18}, 1 (2001).

\bibitem{BWC8}
S. Mukohyama,  Phys. Lett. B {\bf 473}, 241 (2000).

\bibitem{BWC9}
D. Ida, JHEP {\bf 0009}, 014 (2000).

\bibitem{BWC10}
S. Mukohyama, T. Shiromizu, K. Maeda, 
Phys. Rev. {\bf D62}, 024028 (2000).

\bibitem{4D0} J.M. Bardeen, Phys. Rev. {\bf D 22}, 1882  
(1980).

\bibitem{4D1} V.F. Mukhanov, H.A. Feldman, R.H. Brandenberger,  
Phys. Reports, {\bf 215}, 203 (1992).

\bibitem{4D2} 
H. Kodama, M. Sasaki, Int. J. Mod. Phys. {\bf A1}, 265 (1986).

\bibitem{SMS}
T. Shiromizu, K. Maeda, M. Sasaki, 
Phys. Rev. {\bf D62}, 024012 (2000).

\bibitem{II} A. Ishibashi, H. Ishihara, {\it Phys. Rev.} {\bf D56},
3446 (1997).

\bibitem{GT}
J. Garriga. and T. Tanaka, Phys. Rev. Lett. {\bf 84}, 2778 (2000).

\bibitem{TM}
T. Tanaka and X. Montes, Nucl. Phys. {\bf B582}, 259 (2000). 


\bibitem{SSM}
M. Sasaki, T. Shiromizu and K. Maeda,  
Phys. Rev. {\bf D62}, 024008 (2000).

\bibitem{GKR}
S. Giddings, E. Katz and L. Randall, 
JHEP {\bf 0003}, 023 (2000). 

\bibitem{MWBH}
R. Maartens, D. Wands, B. A. Bassett and I. Heard, Phys. Rev 
{\bf D62}, 041301 (2000).

\bibitem{HHR}
S. W. Hawking, T. Hertog, H. S. Reall,
Phys.Rev. {\bf D62}, 043501 (2000).

\bibitem{pre}
S. Kobayashi, K Koyama and J. Soda, Phys. Lett. {\bf B501}, 157 
(2001).

\bibitem{BWCP1} 
H. Kodama, A. Ishibashi, O. Seto, Phys. Rev. {\bf D62},
064022 (2000).

\bibitem{BWCP2}  
R. Maartens, Phys. Rev {\bf D62}, 084023 (2000).

\bibitem{BWCP3}  
D. Langlois, Phys. Rev {\bf D62}, 126012 (2000).

\bibitem{BWCP4} 
C. van de Bruck, M. Dorca, R.H. Brandenberger, A. Lukas, 
Phys. Rev. {\bf D62}, 123515 (2000).

\end{thebibliography}
\end{document}